\documentclass[useAMS]{mn2e}
\usepackage{graphicx}

\def\simgt{\ {\raise-.5ex\hbox{$\buildrel>\over\sim$}}\ }
\def\I{\'\i}
\def\cd{cd$^{-1}$\,}
\def\kms{kms$^{-1}$\,}

\begin{document}

\title[Photometric studies of $\beta$ CMa, 15 CMa and KZ Mus]
{Photometric studies of three multiperiodic $\beta$ Cephei stars: $\beta$ 
CMa, 15 CMa and KZ Mus}
\author[G. Handler et al.]
 {R. R. Shobbrook,$^{1}$ G. Handler,$^{2}$ D. Lorenz,$^{2}$ 
D. Mogorosi\,$^{3}$
\and \\
$^{1}$ Research School of Astronomy and Astrophysics, Australian National
University, Canberra, ACT, Australia\\
$^{2}$ Institut f\"ur Astronomie, Universit\"at Wien,
T\"urkenschanzstrasse 17, A-1180 Wien, Austria\\
$^{3}$ Department of Physics, University of the North-West, Private Bag
X2046, Mmabatho 2735, South Africa
}

\date{Accepted 2005 July 17.
 Received 2005 August 13;
in original form 2005 September 10} 
\maketitle 
\begin{abstract} 

We have carried out single and multi-site photometry of the three $\beta$
Cephei stars $\beta$ and 15 CMa as well as KZ Mus. For the two stars in
CMa, we obtained 270 h of measurement in the Str\"omgren $uvy$ and Johnson
$V$ filters, while 150 h of time-resolved Str\"omgren $uvy$ photometry was
acquired for KZ Mus. All three stars are multi-periodic variables, with
three ($\beta$ CMa) and four (15 CMa, KZ Mus) independent pulsation modes.
Two of the mode frequencies of 15~CMa are new discoveries and one of the
known modes showed amplitude variations over the last 33 years. Taken
together, this explains the star's diverse behaviour reported in the
literature fully.

Mode identification by means of the amplitude ratios in the different
passbands suggests one radial mode for each star. In addition, $\beta$ CMa
has a dominant $\ell=2$ mode while its third mode is nonradial with
unknown $\ell$. The nonradial modes of 15 CMa, which are $\ell \leq 3$,
form an almost equally split triplet that, if physical, would imply that
we see the star under an inclination angle larger than 55\degr. The
strongest nonradial mode of KZ Mus is $\ell=2$, followed by the radial
mode and a dipole mode. Its weakest known mode is nonradial with unknown
$\ell$, confirming previous mode identifications for the star's
pulsations.

The phased light curve for the strongest mode of 15 CMa has a descending
branch steeper than the rising branch. A stillstand phenomenon during the
rise to maximum light is indicated. Given the low photometric amplitude of
this nonradial mode this is at first sight surprising, but it can be
explained by the mode's aspect angle.

\end{abstract}

\begin{keywords}
stars: variables: other -- stars: early-type -- stars: oscillations
-- stars: individual: $\beta$ CMa -- stars: individual: 15 CMa -- 
stars: individual: KZ Mus  -- techniques: photometric
\end{keywords}

\section{Introduction}

The $\beta$ Cephei stars are a group of early B-type stars with masses
between 9 and 17~$M_{\odot}$ that exhibit light, radial velocity and
line-profile variability on time scales between 2--8 hours (Stankov \&
Handler 2005). The cause of their variability is pulsation in pressure and
gravity modes of low radial order. About 110 definite members of this
group are known in our Galaxy (Stankov \& Handler 2005, Pigulski 2005,
Handler 2005), but a large number of candidates was also revealed in the 
Large Magellanic Cloud (Ko{\l}aczkowski et al.\ 2004).

One of the major astrophysical applications of the $\beta$~Cephei stars is
asteroseismology, the determination of the interior structure of pulsating
stars via theoretical modelling of their normal mode spectra. All
main-sequence pulsators more massive than the Sun that have been studied
asteroseismically so far, and where unambiguous results on stellar
structure could be obtained (e.g.\ see Pamyatnykh, Handler \& Dziembowski
2004), belong to the $\beta$ Cephei stars.

The main prerequisite for an asteroseismic study is that a number of
pulsation modes large enough is known to remove any ambiguities concerning
the position of the star in the HR diagram, and that these modes are
unambiguously identified with their pulsational quantum numbers, $k$, the
radial overtone of the mode, the spherical degree $\ell$, and the
azimuthal order $m$. Even among the $\beta$ Cephei stars, such objects are
not easy to find, which can often be traced to insufficient data that are
incapable of providing the required mode frequency resolution and number
of mode detections.

Consequently, it is worthwhile to select potentially promising targets and 
to obtain reasonably large amounts of time-resolved photometric and/or 
spectroscopic data so that a reliable assessment of their asteroseismic 
promise can be made. For the present work, we have chosen three stars that 
seemed to justify such an effort.

The conspicuous naked-eye star ($V=1.98$) $\beta$ CMa is known as a
radial-velocity variable for more than a hundred years (see Struve 1950).
Shobbrook (1973a) obtained the first extensive photometric data set for
this star and analysed it together with the radial velocities previously
published. He found three pulsation periods for the star, and suggested
that all of them changed somewhat over the years. As his measurements were
taken in the $V$ filter only, no photometric mode identification could be
made, but the different light-to-radial-velocity amplitudes of the two
strongest modes noticed by Shobbrook (1973a) suggest that they have
different $\ell$. However, we should note that this conclusion is based on
the assumption that both modes had the same intrinsic amplitude at the
different epochs of the photometric and spectroscopic measurements.

15 CMa ($V=4.80$) shows notoriously difficult variability according to the
literature. This began with the first claim of its radial velocity
variability (Campbell 1911) that was later retracted (Moore 1936). Among
the successive studies of its variability, the photometric investigation
by Shobbrook (1973b) was the most extensive. He identified the period of
the dominant variation in his light curves, and noted its non-sinusoidal
shape, but could not push the analysis further. Heynderickx (1992)
suggested the presence of at least two pulsation modes, which were later
identified with $\ell=2$ and 0, respectively, on the basis of multicolour
photometry (Heynderickx, Waelkens \& Smeyers 1994).

KZ Mus ($V=9.1$) has been studied during a two-site photometric effort by
Handler et al.\ (2003). These authors also review the (short) history of
the investigations of its variability. They found four independent modes
in the stellar light variations, where the strongest of these have
$\ell=2$, 0 and 1, respectively, and the radial mode is either the
fundamental or the first overtone. The presence of additional pulsation
modes was suspected by Handler et al.\ (2003), who also pointed out the
importance of an identification of the spherical degree of the fourth
known mode.

\section{Observations and reductions}

We acquired time-resolved differential photoelectric photometry of all the
three targets. The bright stars $\beta$ and 15~CMa were measured with
respect to the comparison stars HR 2271 ($V=5.8$, B3II/III) and 17 CMa
($V=5.8$, A2V), that were observed alternating with the targets.  No
evidence for variability of the latter two stars within a limit of 
1.5~mmag was found.

This group of four stars was observed from three observatories in the time
span of November 2004 to February 2005. Thirty-five nights of Str\"omgren
$uvy$ data were obtained at Siding Spring Observatory, Australia, with the
0.6-m telescope. Eighteen nights were acquired with the 0.5-m telescope at
the South African Astronomical Observatory. The Johnson $V$ filter was
used for the first five nights; the remaining 13 nights consist of
Str\"omgren $uvy$ data, but the $u$ filter was not used for $\beta$ CMa at
this site.  Finally, ten nights of measurement were performed with the
0.75-m Automated Photometric Telescope T6 at Fairborn Observatory, USA,
again using Str\"omgren $uvy$ filters. Neutral density filters were used
in all cases to avoid photomultiplier destruction, owing to the extreme
brightness of $\beta$ CMa; the same neutral density filters were applied
to all stars at each site. The total of 63 nights of observation has a
time base of 111.95~d, corresponding to 272.7 h of measurements of $\beta$
CMa and 265.4 h of measurements for 15 CMa, respectively.

KZ Mus was measured with the same observational technique. The comparison
stars were HD 110022 ($V=7.8$, B8 III/IV), which was replaced by HD 109082
($V=8.1$, B9IV) after the first few nights when we found evidence for
variability of the former object. The second comparison star was HD 111876
($V=7.1$, B9.5/A0V). HD 109082 was constant during the observations.
However, HD 111876 showed a small ($\sim$ 0.01 mag in all three filters)  
mean magnitude drift during the total time span of observations.

This group of stars was observed from March to June 2005, but with the
0.6-m telescope at Siding Spring Observatory only. Data were acquired
during 27 nights, giving a total time span of 98.9 d, and 154.3 h of
measurement resulted. Again the Str\"omgren $uvy$ filters were used, but
no neutral density filters had to be employed.

Data reduction comprised a correction for coincidence losses, subtraction
of sky background and extinction determination and correction. Whenever
possible, nightly extinction coefficients were determined with the
classical Bouguer method (fitting a straight line to a plot of magnitude
vs.\ air mass) from the measurements of the two comparison stars. For the 
remaining nights, mean coefficients from the adjoining nights were used.

The magnitudes of the comparison stars were adjusted to their mean
differences, and were then combined into a curve that was assumed to
reflect the effects of sky transparency and detector sensitivity changes
only. Subsequently, these combined time series were binned into intervals
that would allow good compensation for the above-mentioned non-intrinsic
variations in the target star time series and were subtracted from the
measurements of the variables. The binning minimises the introduction of
noise in the differential light curves of the targets.

The timings for the differential light curves were heliocentrically
corrected as the next step. Finally, the photometric zeropoints of the
different instruments were compared between the different sites and were
adjusted if necessary. Measurements in the Str\"omgren $y$ and Johnson $V$
filters were treated as equivalent due to the similar effective
wavelengths of these filters, and were analysed jointly after a correction
for differential colour extinction in $V$. Some example light curves of
our three targets together with multifrequency solutions to be discussed
in the following are shown in Figs.\ 1, 2 and 3, respectively.

\begin{figure}
\includegraphics[width=88mm,viewport=00 00 318 450]{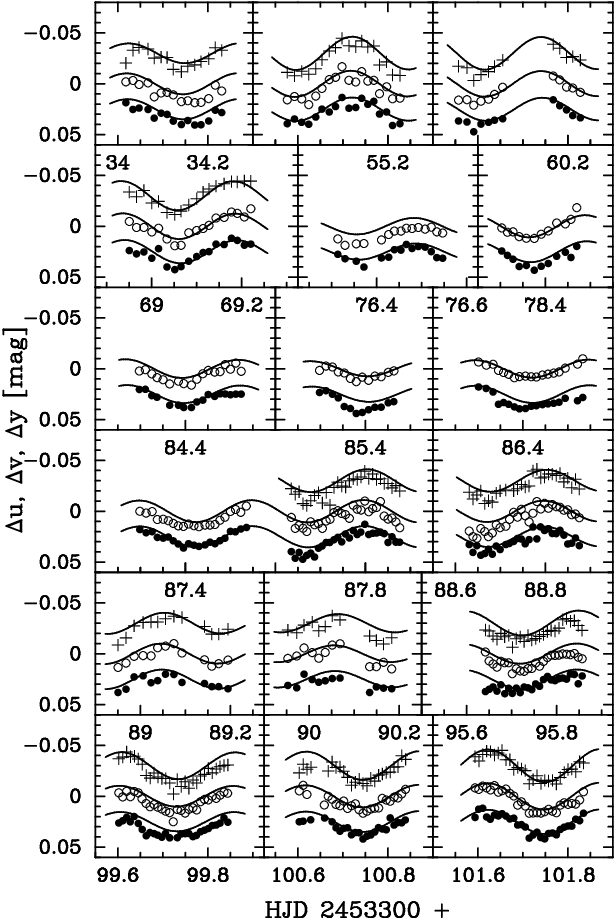}
\caption[]{Some of our observed light curves of $\beta$ CMa. Plus signs
are data in the Str\"omgren $u$ filter, opem circles are our $v$
measurements and filled circles represent Str\"omgren $y$ data. The full
line is a fit composed of all the periodicities detected in the light
curves (Table 2). The amount of data displayed is approximately one third 
of the total.}
\end{figure}

\begin{figure}
\includegraphics[width=88mm,viewport=00 00 318 450]{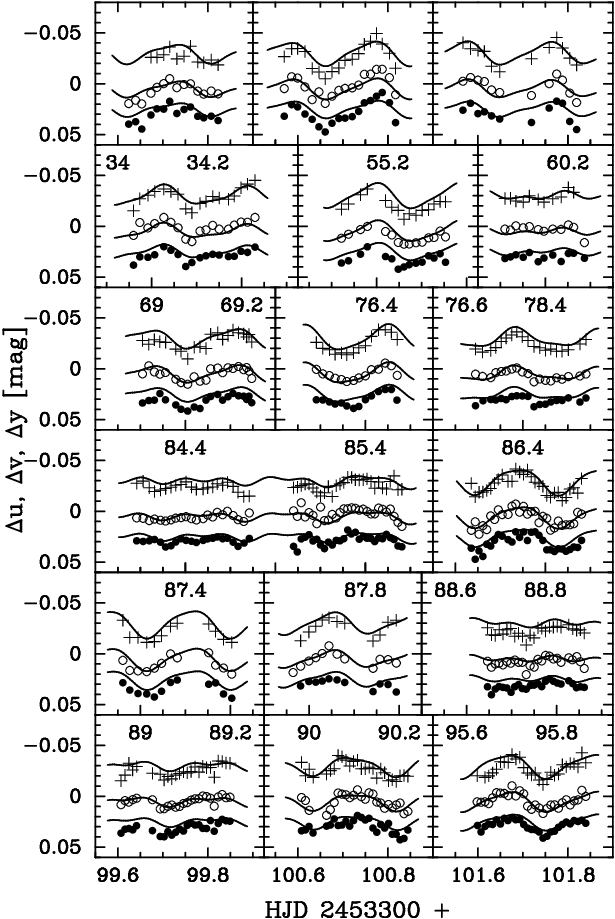}
\caption[]{Some of our observed light curves of 15 CMa. Plus signs are
data in the Str\"omgren $u$ filter, open circles are our $v$
measurements and filled circles represent Str\"omgren $y$ data. The full
line is a fit composed of all the periodicities detected in the light
curves (Table 3). The amount of data displayed is approximately one third 
of the total.}
\end{figure}

\begin{figure}
\includegraphics[width=88mm,viewport=00 00 322 315]{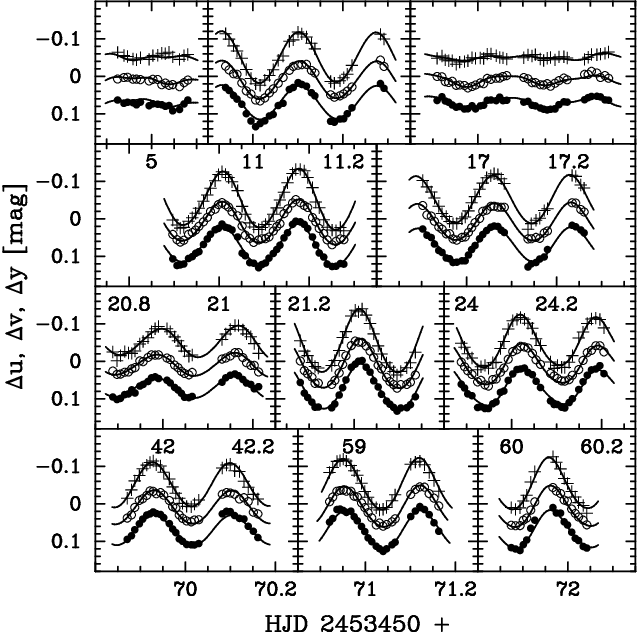}
\caption[]{Some of our observed light curves of KZ Mus. Plus signs are
data in the Str\"omgren $u$ filter, opem circles are our $v$
measurements and filled circles represent Str\"omgren $y$ data. The full
line is a fit composed of all the periodicities detected in the light
curves (Table 4). The amount of data displayed is approximately half the 
total.}
\end{figure}

\section{Frequency analysis}

Our frequency analysis was mainly performed with the program {\tt
Period98} (Sperl 1998). This package applies single-frequency power
spectrum analysis and simultaneous multi-frequency sine-wave fitting. It
also includes advanced options such as the calculation of optimal
light-curve fits for multi-periodic signals including harmonic,
combination, and equally spaced frequencies. Our analysis will require 
some of these features.

\subsection{$\beta$ CMa}

We started by computing the Fourier spectral window of the $u$ filter
data. It was calculated as the Fourier transform of a single noise-free
sinusoid with a frequency of 3.979 \cd (the strongest pulsational signal
of $\beta$ CMa) and an amplitude of 12 mmag, sampled in the same way as
our measurements. The upper panel of Fig.\ 4 contains the result. The
alias structures in this window are sufficiently suppressed to allow easy
and unique determinations of the frequencies of the stellar light
variations.

\begin{figure}
\includegraphics[width=88mm,viewport=-03 05 245 340]{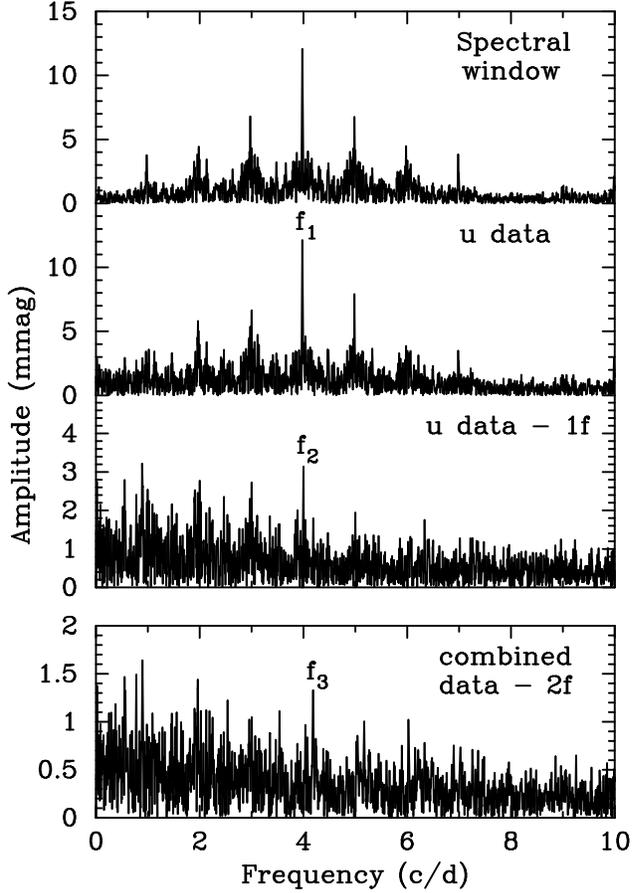}
\caption[]{Amplitude spectra of $\beta$ CMa. The uppermost panel shows the 
spectral window of the data, followed by the periodogram of the data. 
Successive prewhitening steps are shown in the following panels; note 
their different ordinate scales.}
\end{figure}

We proceeded to compute the amplitude spectra of the data themselves
(second panel of Fig.\ 4). The signal designated $f_1$ dominates. We
prewhitened it by subtracting a synthetic sinusoidal light curve with a
frequency, amplitude and phase that yielded the smallest possible residual
variance. We then computed the amplitude spectrum of the residual light
curve, which is shown in the third panel of Fig.\ 4.

A second signal ($f_2$) can clearly be seen in this graph. Prewhitening it
leaves no further significant periodicities in the $u$ data alone. At this
stage, we prewhitened the data from the $v$ and $y$ filters by the $f_1$
and $f_2$ frequencies, then combined these residuals with those from the
$u$ filter data divided by 1.2. This 1.2 factor scales the $u$ data to the
same amplitude as that found for $f_1$ in the $v$ and $y$ data. This
procedure is valid because the light curves of $\beta$ Cephei stars have
the same phase at all optical wavelengths.

The residual amplitude spectrum of the $uvy$ data combined in this way is
shown in the lowest panel of Fig.\ 4. We note the presence of another
signal, $f_3$, and include it in our frequency solution for $\beta$ CMa.
Prewhitening by this variation as well, we find no further significant
periodicity in our combined residual amplitude spectrum.

We consider an independent peak statistically significant if it exceeds an
amplitude signal-to-noise ratio of 4 in the periodogram; combination
signals must satisfy $S/N>3.5$ to be regarded as significant (see Breger
et al.\ 1993, 1999 for a more in-depth discussion of this criterion). The
noise level was calculated as the average amplitude in a 5~\cd interval
centred on the frequency of interest.

We are now in a position to determine a final multifrequency solution for
our light curves of $\beta$ CMa. To this end, we first computed a weighted
average for the frequencies derived from the individual filters' data,
where the weight corresponded to the $S/N$ of the signals. The results
were adopted as our final frequency values. We then fitted these
frequencies to the $uvy$ data and determined their amplitudes and phases
(that were, as assumed previously, found to be the same for all filters
within the errors). The final solution is listed in the upper part of
Table 1.

\begin{table}
\caption[]{Multifrequency solutions for photometric data of $\beta$ CMa. 
Error estimates (following Montgomery \& O'Donoghue 1999) on the 
amplitudes are $\pm$ 0.4 mmag in $u$ and $\pm$ 0.3 mmag in $v$ and $y$, 
respectively. The S/N ratio was computed following Breger et al.\ (1993) 
and is quoted for the $v$ filter.}
\begin{center}
\begin{tabular}{lccccc}
\hline
ID & Freq. & $u$ Amp. & $v$ Amp. & $y$ Amp. & $S/N$ \\
 & (\cd) & (mmag) & (mmag) & (mmag) & \\
\hline
\multicolumn{6}{c}{New photometry}\\
\hline
$f_{1}$ & 3.9793 $\pm$ 0.0001 & 12.0 & 10.6 & 10.0 & 25.5 \\ 
$f_{2}$ & 3.9994 $\pm$ 0.0007 & 3.8 & 1.7 & 1.3 & 4.1 \\
$f_{3}$ & 4.1857 $\pm$ 0.0007 & 1.7 & 1.7 & 1.4 & 4.2 \\ 
\hline
\multicolumn{6}{c}{Archival data}\\
\hline
$f_{1}$ & 3.9792 $\pm$ 0.0001 & & & 10.3 $(V)$ & 28.5 \\
$f_{2}$ & 3.9999 $\pm$ 0.0007 & & & 1.9 $(V)$ & 5.2 \\
$f_{3}$ & 4.1821 $\pm$ 0.0009 & & & 1.4 $(V)$ & 4.0 \\
\hline
\end{tabular}
\end{center}
\end{table}

The error estimates we give are the formal values for the amplitudes
(Montgomery \& O'Donoghue 1999). We note that such formal error bars are
expected to underestimate the real errors by about a factor of 2 (Handler
et al.\ 2000, Jerzykiewicz et al.\ 2005). For the frequencies themselves,
we computed both the formal errors, and the deviation of the mean of the
values from the individual filters; we then adopted the larger of the two
values. For frequencies $f_1$ and $f_3$ the two results agreed quite well
(within 10 per cent). However, the error of the mean was a factor of 2
larger for~$f_2$.

This result can be easily understood: within the time span of our data,
$f_2$ is not fully resolved from a possible variation of 4 cycles per
sidereal day. Such a variation could be present in our light curves due to
residual (colour) extinction effects, despite our careful data reduction.  
This interpretation is supported by the observation that the value of
$f_2$ is closest to 4 cycles per sidereal day in the $u$ band where
possible residual extinction effects are expected to be largest. We must
therefore be careful in the interpretation of the frequency and $uvy$
amplitudes of $f_2$.

\subsubsection{Reanalysis of archival data}

One of us (RRS) has digitised his archival $V$ filter photometry of
$\beta$ and 15 CMa taken in the early 1970s. We can therefore analyse
these measurements with our frequency analysis methods, allowing a direct
comparison of the results.

For $\beta$ CMa we confirm the findings by Shobbrook (1973a, see lower
part of Table 1): we obtain the same three frequencies as he did. Given
the probable underestimation of the real errors of the frequencies and
amplitudes (especially for mode $f_2$), we are reluctant to claim the
presence of amplitude and/or frequency variability for the star.

On the other hand, we note that there seems to be some low-frequency
variability, for which we are unable to determine a period, remaining in
the residuals, similar to what can be discerned in the lowest panel of
Fig.\ 4. We find this behaviour in all independent data sets we have for
$\beta$ CMa, but not in the differential light curve of the two comparison
stars. Because we can exclude an extinction effect for its origin, we are
inclined to believe that this variation is intrinsic to our target star.

We attempted to obtain an improved frequency solution and to check for
amplitude and frequency variability of the stellar pulsation modes by
analysing the combined visual photometric data by Shobbrook (1973a),
Balona et al.\ (1996) and the HIPPARCOS satellite (ESA 1997) together with
our measurements. Regrettably, the amount and temporal separation of the
data render this investigation inconclusive: we derive a best-fitting
frequency for the strongest mode of $3.9793334 \pm 0.0000009$\,\cd that is
slightly variable in amplitude and phase, but we cannot be certain whether
this is a numerical artifact due to beating with the close mode at
3.9994\,\cd when unresolved in certain subsets of data.

\subsection{15 CMa}

The frequency analysis for this star was carried out in a similar way as
that for $\beta$ CMa and is illustrated in Fig.\ 5. This time we use the
$v$ data for display as they resulted in the light curves of best $S/N$.
In the case of 15 CMa, the spectral window had a single noise-free
sinusoid with a frequency of 5.419 \cd and an amplitude of 5.4 mmag as
input (upper panel of Fig.\ 5). The amplitude spectrum of the data
themselves (second panel)  already makes it clear that more than one
frequency is present in the light curves. Prewhitening the strongest
signal ($f_1$) leaves two conspicuous peaks in the residual amplitude
spectrum. One of these is the first harmonic of $f_1$. Further
prewhitening with these three signals indicates the presence of another
periodicity, and prewhitening that one as well reveals a fourth
independent frequency in the light curves. Pushing the analysis further,
also by combining the residuals for the $u$, $v$ and $y$ data, as we did
for $\beta$ CMa, did not result in the detection of additional
periodicities.

\begin{figure}
\includegraphics[width=88mm,viewport=-03 05 245 395]{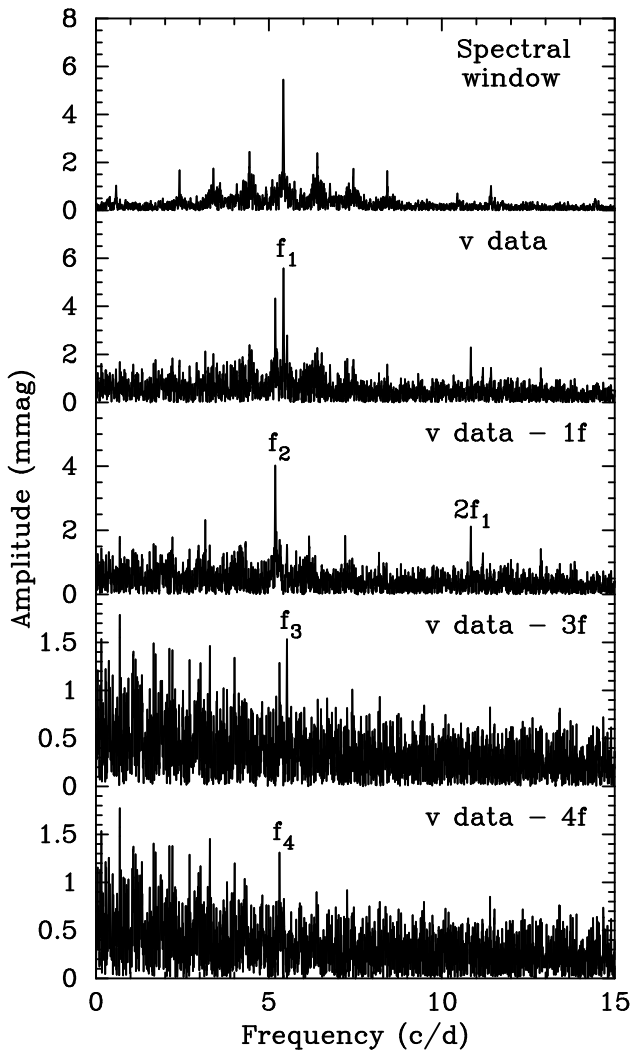}
\caption[]{Amplitude spectra of 15 CMa. The uppermost panel shows the 
spectral window of the data, followed by the periodogram of the data. 
Successive prewhitening steps are shown in the following panels; note 
their different ordinate scales.}
\end{figure}

The final multifrequency solution for 15 CMa was determined in the same
way as for $\beta$ CMa: by computing $S/N$-weighted averages of the
frequencies determined from the individual filters' data. We note that,
for reasons of better accuracy, we fixed the frequency of the harmonic
signal to exactly twice the value of the parent mode with {\tt Period 98}.

We then recomputed the $uvy$ amplitudes with these ``best'' frequencies
and list the result in Table 2. Again, the adopted error estimates on the
frequencies correspond to the larger of the two values derived from
analytic formulae and rms deviations of the mean of the frequencies.

\begin{table}
\caption[]{Multifrequency solution for time-resolved photometric data of
15 CMa. Error estimates (following Montgomery \& O'Donoghue 1999) on the
amplitudes are $\pm$ 0.3 mmag in $u$ and $\pm$ 0.2 mmag in $v$ and $y$,
respectively. The S/N ratio was computed following Breger et al.\ (1993,
1999) and is quoted for the $v$ data.}
\begin{center}
\begin{tabular}{lccccc}
\hline
ID & Freq. & $u$ Amp. & $v$ Amp. & $y$ Amp. & $S/N$ \\
 & (\cd) & (mmag) & (mmag) & (mmag) & \\
\hline
\multicolumn{6}{c}{New photometry}\\
\hline
$f_{1}$ & 5.4187 $\pm$ 0.0002 & 6.2 & 5.4 & 4.6 & 16.0 \\
$f_{2}$ & 5.1831 $\pm$ 0.0003 & 5.4 & 3.9 & 2.9 & 11.2 \\
2$f_{1}$ & 10.8374      &  2.4  & 2.1 &     2.1 &  7.2 \\
$f_{3}$ & 5.5212 $\pm$ 0.0008 & 2.0 & 1.6 & 1.3 &  4.7 \\
$f_{4}$ & 5.3085 $\pm$ 0.0012 & 1.6 & 1.4 & 1.1 &  4.2 \\
\hline
\multicolumn{6}{c}{Archival data}\\
\hline
$f_{1}$ & 5.4180 $\pm$ 0.0003 & & & 4.8 (V) & 12.3\\
2$f_{1}$ & 10.8359 & & & 1.9 (V)& 6.2\\
$f_2$ & 5.1845 $\pm$ 0.0009 & & & 1.5 (V) & 3.8\\
$f_3$ & 5.5253 $\pm$ 0.0007 & & & 1.9 (V) & 4.9\\
$f_4$ & 5.3103 $\pm$ 0.0008 & & & 1.6 (V) & 4.3\\
\hline
\end{tabular}
\end{center}
\end{table}

\subsubsection{Reanalysis of archival data}

We analysed the $V$ measurements of 15 CMa by Shobbrook (1973b, also
digitised by RRS), who also detected our signal $f_1$ and its first
harmonic. However, he could not continue to search for further frequencies
because the residual amplitude spectrum was too complicated.

Thanks to our multi-site measurements, we know four pulsation frequencies
of the star without aliasing ambiguities, and we can use this information
in our reanalysis. This helps us to solve the puzzle that the data by
Shobbrook (1973b) provided: we first detect $f_1$ and its harmonic. The
three other frequencies detected in our new measurements are also present
in the older photometry, and all three are required to explain the
observations, as proven by different prewhitening trials. We list the 
result of this analysis in the lower part of Table 2. We point out that 
$f_2$ has a $S/N$ below 4, but since it is known to be an intrinsic signal 
from our new measurements, we only require $S/N>3.5$ to accept this 
frequency as real in the old data.

In this case, $f_2$ can be safely claimed as having varied in amplitude,
whereas no such statement can be made for the other signals. In this
context it is interesting to revisit the discussion of the Geneva data by
Heynderickx (1992), who also observed 15 CMa. He detected our $f_1$ and
its harmonic, but then encountered two closely spaced signals around
$f_2$, where the frequency difference of these signals is close to the
inverse time span of his data set. Consequently, it can be suspected that
these two apparent signals near $f_2$ could in fact be due to a single
mode with amplitude variability during Heynderickx' observations. We note
that an attempt to refine the pulsation frequencies of the star by
analysing the measurements by Shobbrook (1973b) and the HIPPARCOS
satellite (ESA 1997) together with ours did not bear fruit due to aliasing
problems.

We also note that, just as for $\beta$ CMa, some slow aperiodic
variability seems to be left in the residuals of both the archival and the
new measurements. This slow variability is not correlated with the one of
$\beta$ CMa (neither in the old nor in the new data), which is further
support for the idea that it is intrinsic to the stars and is not an
instrumental effect.

\subsubsection{The light curve shape of $f_1$}

The amplitude of the harmonic frequency of $f_1$ is unusually high given 
the amplitude of the independent mode to which it is connected. We have 
therefore examined the shape of the light curve due to this mode. We 
fitted the five known frequencies of 15 CMa to the archival data and to 
our new measurements and then prewhitened the fit composed of the 
resulting parameters for $f_2$, $f_3$ and $f_4$. We then phased the 
residuals with $f_1$ and show the results in Fig.\ 6.

\begin{figure}
\includegraphics[width=88mm,viewport=00 00 268 316]{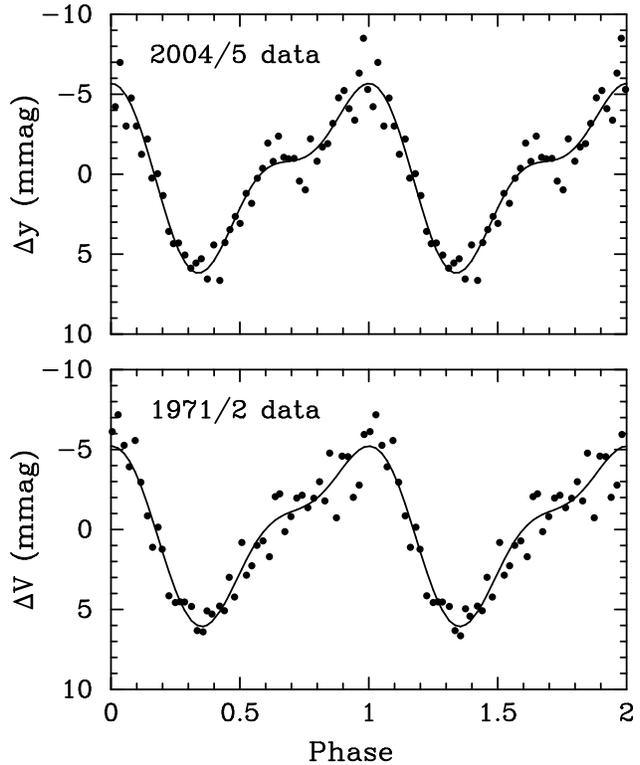}
\caption[]{Phase diagram of light curves of 15 CMa with respect to $f_1$
(full dots), compared to a fit computed with $f_1$ and 2$f_1$. Two cycles
of the variation are shown. The rising branch of this light variation is
flatter as the descending branch, and there seems evidence for a
stillstand before light maximum.}
\end{figure}

The phase diagrams obtained in this way appear odd for a pulsating star:  
the rising branch of the light curve is less steep than the descending
branch. When computing a fit to these phase diagrams composed of $f_1$ and
2$f_1$, or when averaging the two phase diagrams, it even seems possible
that there is a ``stillstand'' phenomenon in the rising branch. We would
however like to see more data on the star before coming to a definite
conclusion on this matter.

It is interesting to note that the scatter in the two phase diagrams are
very similar: although smaller in number, the precision of the
measurements taken in the 1970s is slightly better than that of our
present data.

\subsection{KZ Mus}

The frequency analysis for this star was carried out in a similar fashion
as for the others. We determined individual frequencies step-by-step and
found four independent modes (see Fig.\ 7) plus three combination
frequencies. Again, the frequencies were first determined for each
individual filter, using {\tt Period 98} to fix the frequencies of
combination signals to the exact sums predicted by those of parent modes.  
Then, $S/N$-weighted averages of the frequencies determined from the
individual filters' data were computed and adopted as our final frequency 
values. Our multifrequency solution obtained this way is listed in Table 
3.

\begin{figure}
\includegraphics[width=88mm,viewport=-03 05 245 537]{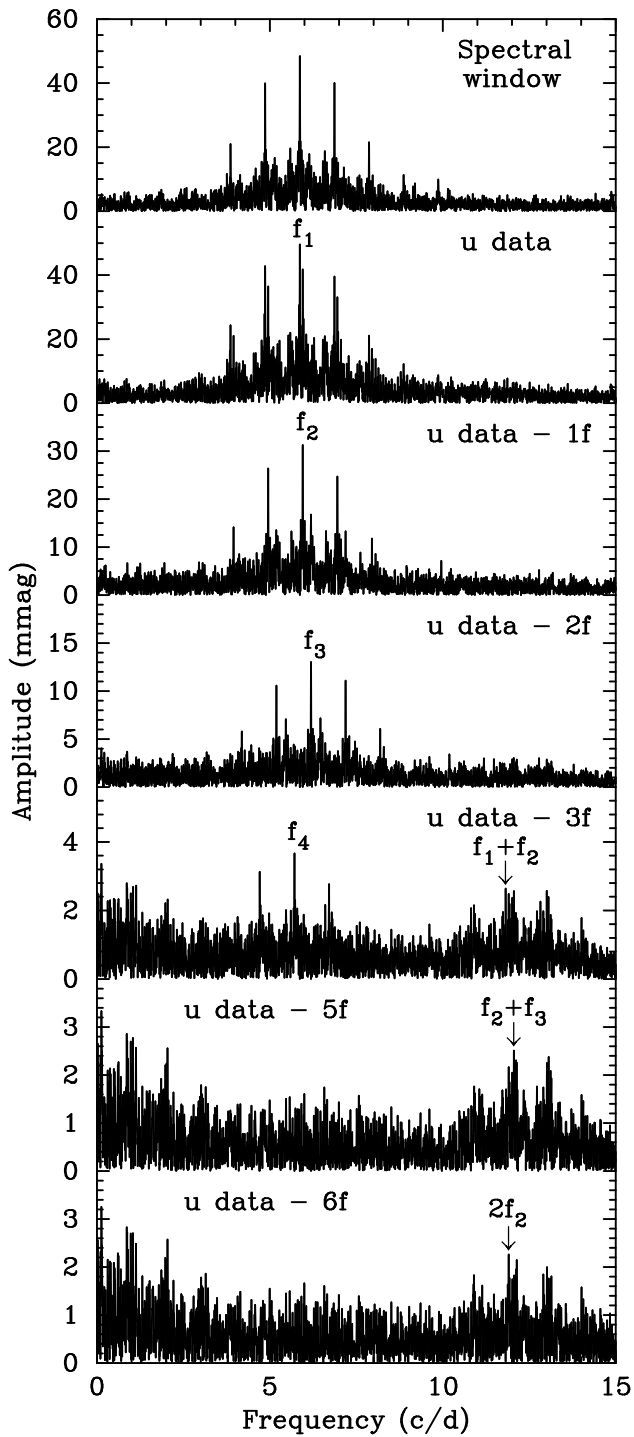}
\caption[]{Amplitude spectra of KZ Mus. The uppermost panel shows the 
spectral window of the data, followed by the periodogram of the data. 
Successive prewhitening steps are shown in the following panels; note 
their different ordinate scales.}
\end{figure}

\begin{table}
\caption[]{Multifrequency solution for our new time-resolved photometry of
KZ Mus. Error estimates (following Montgomery \& O'Donoghue 1999) on the
amplitudes are $\pm$ 0.4 mmag in $u$ and $v$, and $\pm$ 0.3 mmag in $y$.
The S/N ratio was computed following Breger et al.\ (1993, 1999) and is
quoted for the $u$ filter data.}
\begin{center}
\scriptsize
\begin{tabular}{lccccc}
\hline
ID & Freq. & $u$ Amp. & $v$ Amp. & $y$ Amp. & $S/N$ \\
 & (\cd) & (mmag) & (mmag) & (mmag) & \\
\hline
$f_1$ & 5.86400 $\pm$  0.00004 & 47.2  &  41.0 &   38.4 &   85.4\\
$f_2$ & 5.95059 $\pm$  0.00007 & 35.0  &  20.6 &   17.3 &   63.4\\
$f_3$ & 6.18774 $\pm$  0.00013 & 16.1  &  12.4 &   11.6 &   29.3\\
$f_4$ & 5.7094  $\pm$  0.0005  & 4.0   & 3.2  &  3.6  &  7.1\\
$f_1+f_2$ & 11.81459           & 1.7   &  1.7 &  1.7  &   3.2\\
$f_1+f_3$ & 12.05174           & 2.3   &  1.8 &  1.7  &   4.3\\
2$f_2$ &    11.90118           & 2.2   &  1.6 &  1.8  &   4.1\\
\hline
\end{tabular}
\normalsize
\end{center}
\end{table}

Comparing our results to those by Handler et al.\ (2003), we cannot find
significant evidence for amplitude or frequency variability, but we do
note the occurrence of a new combination frequency, $f_1+f_3$. This signal
is actually present in the older measurements, but we were not convinced
about its reality during the previous analysis.

On the other hand, the combination signal $f_2-f_1$ detected by Handler et
al.\ (2003) is not present in our new measurements. This is likely due to
the increased low-frequency noise in the new data caused by the
variability of the comparison star HD 111876, which we were not able to
remove satisfactorily.

\subsubsection{Refining the determination of the pulsation frequencies}

Although the photometry by Handler et al.\ (2003) was taken three years
before the measurements presented in this paper, the frequency
determinations of both works are accurate enough that we can link the
previous $V$ and the present $y$ data to obtain more accurate pulsation
frequencies. The effective wavelengths of these two bands are also
sufficiently similar to render such an approach valid.

Consequently, we performed a joint frequency analysis with {\tt Period
98}, where we could also reveal the presence of one more combination
signal. The result is given in Table 4. We have confirmed that the
frequencies of the three strongest modes and their combination frequencies
are alias-free by including the HIPPARCOS photometry of the star (ESA
1997) as well. We could not make a similar check for mode $f_4$ due to its
low amplitude. In any case, the residual amplitude spectrum after
prewhitening our joint solution contains no evidence that measurable
amplitude or frequency variations have occurred within the 3.4-y time base
of our observations. We have also determined the Str\"omgren $v$
amplitudes given this frequency solution for the combined 2002 and 2005
data. However, we could not do the same for the Johnson $U$ and
Str\"omgren $u$ bands because their wavelength responses are sufficiently
different that the calculated amplitudes of the strongest modes do not
correspond within the errors.

\begin{table}
\caption[]{Multifrequency solution for the combined ground-based
time-resolved $V$ and $y$, as well as $v$ photometry of KZ Mus. The error 
estimates (Montgomery \& O'Donoghue 1999) on the amplitudes are $\pm$ 0.2 
mmag. The S/N ratio was computed following Breger et al.\ (1993, 1999).}
\begin{center}
\scriptsize
\begin{tabular}{lcccc}
\hline
ID & Frequency & $v$ Amp.\ & $V$ or $y$ Amp.\ & $S/N$ \\
 & (\cd) & (mmag) & (mmag) & \\
\hline
$f_1$ & 5.864016 $\pm$  0.000005 & 41.2 & 38.4 &  109.5\\
$f_2$ & 5.950693 $\pm$  0.000011 & 20.4 & 16.5 &   47.2\\
$f_3$ & 6.187472 $\pm$  0.000016 & 12.3 & 10.9 &   31.4\\
$f_4$ & 5.70935  $\pm$  0.00006  & 3.1 & 3.0  &   8.4\\
$f_1+f_2$ & 11.814709           & 1.9 & 1.9  &   6.2\\
2$f_2$ &    11.901386          & 1.5 & 1.5  &   5.0\\
$f_1+f_3$ & 12.051489          & 1.3 & 1.3  &   4.4\\
$f_2+f_3$ & 12.138165 &  1.2 & 1.2 & 4.0\\
\hline
\end{tabular}
\normalsize
\end{center}
\end{table}

\section{Mode identification}

The spherical degree $\ell$ of the pulsation modes of our three target
stars can be identified by comparing their observed $uvy$ amplitudes with
those predicted by theoretical models. Required input for these models are
the mass and effective temperature of the stars, from which their
luminosity obviously follows as well.

For KZ Mus, we simply adopt the values given by Handler et al.\ (2003),
who derived $T_{\rm eff}=26000\pm700$~K and log~$L/L_{\odot}=4.22\pm0.20$.
For the two stars in Canis Major, the calibrations by Crawford (1978)  
applied to the standard Str\"omgren photometry by Crawford, Barnes \&
Golson (1970) results in $E(b-y)=0.027$ for $\beta$~CMa and $E(b-y)=0.031$
for 15 CMa, respectively. The Str\"omgren system calibration by
Napiwotzki, Sch\"onberner \& Wenske (1993) then yields $T_{\rm
eff}=25600\pm1000$~K for $\beta$~CMa and $T_{\rm eff}=26300\pm1000$~K for
15 CMa.

With the Geneva colour indices of the two stars obtained from the Lausanne
Photometric data base, the calibrations by K\"unzli et al.\ (1997) provide
$T_{\rm eff}=25100\pm1100$~K and log $g=3.3\pm0.6$ for $\beta$ CMa, and
$T_{\rm eff}=26000\pm1500$~K and log $g=3.8\pm0.6$ for 15 CMa.

The analysis of IUE spectra of a number of $\beta$ Cephei stars by
Niemczura \& Daszy{\'n}ska-Daszkiewicz (2005) results in $T_{\rm
eff}=24700\pm800$~K and log $g$=3.74 for $\beta$ CMa and $T_{\rm
eff}=24600\pm1000$~K and log $g$=3.81 for 15 CMa.

The HIPPARCOS parallax of $\beta$ CMa is $6.53 \pm 0.66$ mas. With
$V$=1.98 and the reddening as determined before, we obtain $M_v = -4.1 \pm
0.2$. The parallax of 15 CMa ($2.02 \pm 0.70$ mas) is too inaccurate to be
useful. However, 15 CMa is a member of the young open cluster Collinder
121, whose distance was determined as $592\pm28$ pc (de Zeeuw et al.\
1999), which is in reasonable agreement with the usually quoted distance
to the cluster in the pre-HIPPARCOS era, 630 pc (Feinstein 1967). Using
the most recent distance to Collinder 121, we determine $M_v = -4.2 \pm
0.2$ for this star, given that $V$=4.80 and again adopting the amount of
reddening suggested by Str\"omgren photometry.

Summarising the temperature determinations quoted before, we arrive at
final values of $T_{\rm eff}=25100\pm1200$~K for $\beta$ CMa and
$25600\pm1200$~K for 15 CMa. According to Flower (1996), these effective
temperatures correspond to bolometric corrections of $BC=-2.2\pm0.3$~mag
and $BC=-2.3\pm0.4$~mag, respectively. Consequently, the bolometric
luminosities of the two stars are $-6.3\pm0.4$~mag ($\beta$ CMa) and
$-6.5\pm0.4$~mag (15 CMa), respectively.

\begin{figure}
\includegraphics[width=99mm,viewport=5 00 305 260]{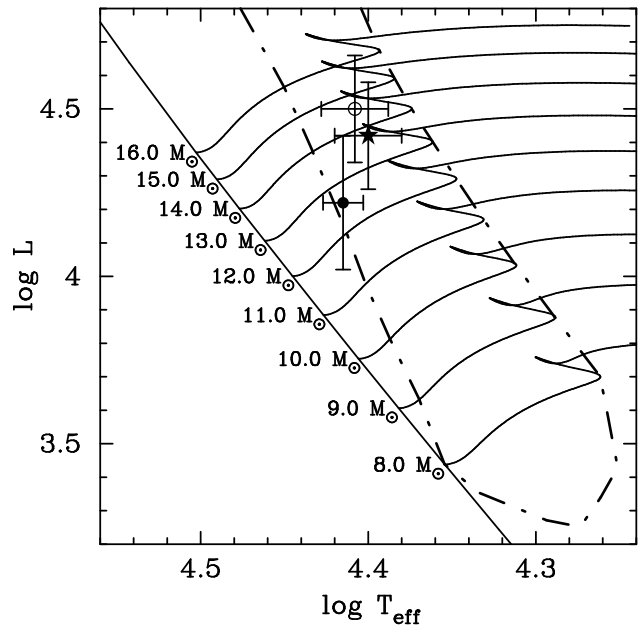}
\caption[]{The positions of our three targets in the theoretical HR
diagram. The filled circle denotes KZ Mus, the star symbol is for
$\beta$~CMa, and the open circle represents 15 CMa. Error estimates on the
effective temperatures and luminosities of the stars are indicated. Some
stellar evolutionary tracks, for a metal abundance of $Z=0.02$, labelled
with their masses (full lines) are included for comparison. We also show
the theoretical borders of the $\beta$~Cephei instability strip
(Pamyatnykh 1999, dashed-dotted line).}
\end{figure}

Figure 8 shows the positions of our three targets in a theoretical HR
diagram. The two stars in Canis Major have very similar properties there,
indicating that both are massive and are approaching the end of their main
sequence life. This is somewhat in contrast with their distinctly
different pulsation periods. KZ Mus appears to just have entered the
instability strip and has lower mass than the two other pulsators.

For the purpose of mode identification we computed theoretical photometric
$uvy$ amplitudes for the $0 \leq \ell \leq 4$ modes of models with masses
between 13.0 and 16.0 $M_{\sun}$ in steps of 1.0$M_{\sun}$, and in a
temperature range of $4.38 \leq \log T_{\rm eff} \leq 4.43$ for $\beta$
and 15 CMa. For KZ Mus, we investigated a range of 11.5 to 14.0
$M_{\sun}$ in steps of 0.5 $M_{\sun}$, and $4.395 \leq \log T_{\rm eff}
\leq 4.435$, therefore applying a generous interpretation of the derived
error bars. All models had an assumed metallicity of $Z=0.02$. This
approach is similar to that by Balona \& Evers (1999).

Theoretical mode frequencies between 3.7 and 4.5 \cd were considered for
$\beta$ CMa, between 4.8 and 5.9 \cd for 15 CMa and between 5.4 and 6.5
\cd for KZ Mus. The frequency ranges chosen are somewhat larger than the
ones excited in the stars to allow for some rotational $m$-mode splitting.
We compare the theoretical photometric amplitude ratios to the observed
ones in Figs.\ 9 to 11 for all independent modes of all three stars.

\begin{figure}
\includegraphics[width=88mm,viewport=00 00 265 455]{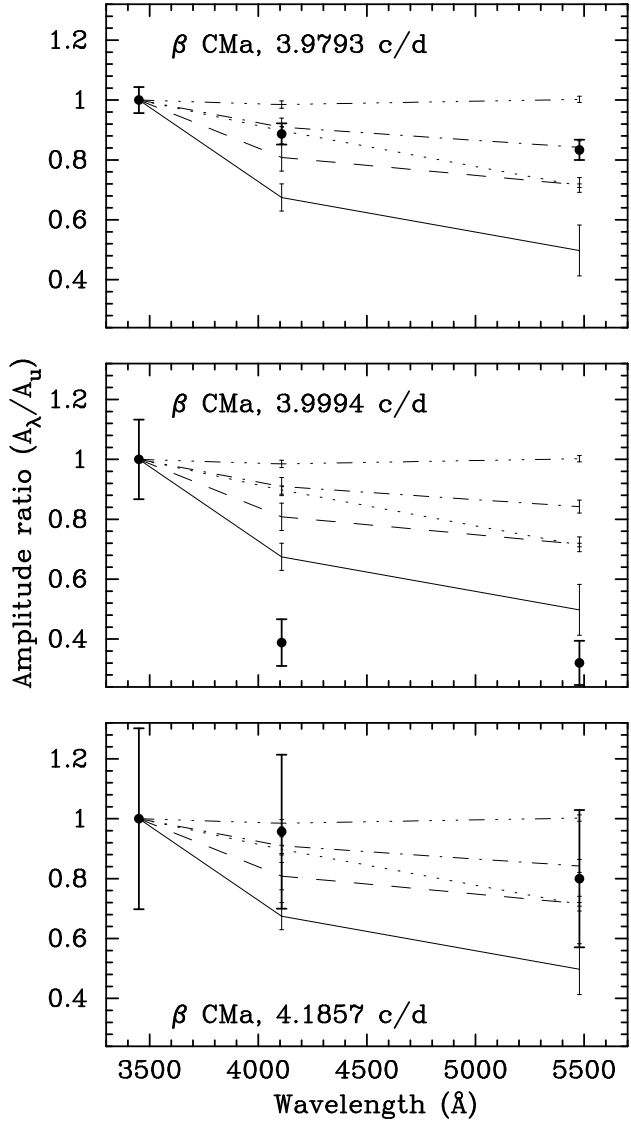}
\caption[]{Mode identifications for $\beta$ CMa from a comparison of
observed and theoretical $uvy$ amplitude ratios, normalised to unity at
$u$. The filled circles with error bars are the observed amplitude ratios.
The full lines are theoretical predictions for radial modes, the dashed
lines for dipole modes, the dashed-dotted lines for quadrupole modes, the
dotted lines for octupole ($\ell=3$) modes, and the dashed-dot-dot-dotted
lines are for $\ell=4$. The thin error bars denote the uncertainties in
the theoretical amplitude ratios.}
\end{figure}

Starting with $\beta$ CMa, it is clear that its strongest mode has a
spherical degree of $\ell=2$. The second mode seems radial, although its 
measured amplitude vs.\ wavelength dependence is considerably steeper than 
theoretically predicted. However, we believe that this is caused by 
problems with differential colour extinction, which artificially increases 
the measured $u$ amplitude, as discussed in Sect.\ 3. In any case, no 
mode identification other than $\ell=0$ is possible for this mode. The 
lowest-amplitude pulsation mode is nonradial, but its spherical degree 
cannot be constrained due to its small amplitude.

\begin{figure}
\includegraphics[width=88mm,viewport=00 00 265 600]{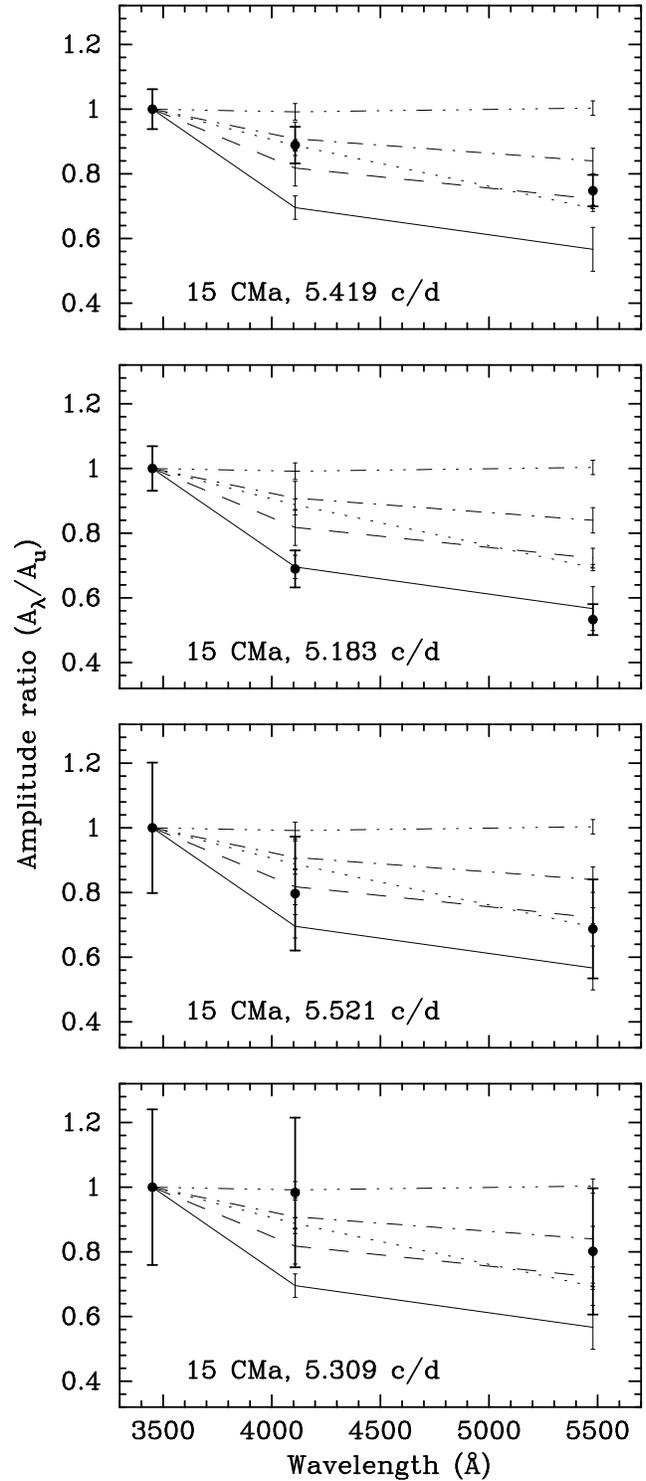}
\caption[]{Mode identifications for 15 CMa from a comparison of observed
and theoretical $uvy$ amplitude ratios, normalised to unity at $u$. The
filled circles with error bars are the observed amplitude ratios. The full
lines are theoretical predictions for radial modes, the dashed lines for
dipole modes, the dashed-dotted lines for quadrupole modes, the dotted
lines for $\ell=3$ modes, and the dashed-dot-dot-dotted lines are for
$\ell=4$. The thin error bars denote the uncertainties in the theoretical
amplitude ratios.}
\end{figure}

Continuing with 15 CMa, we see that the strongest mode is nonradial with
$\ell=1$, 2 or 3, but no distinction between these hypotheses can be
made. The second strongest mode is most likely radial, and no mode
identification is possible for the two weakest modes. However, due to
their proximity to the radial mode in frequency, they must be nonradial.

\begin{figure}
\includegraphics[width=88mm,viewport=00 00 265 600]{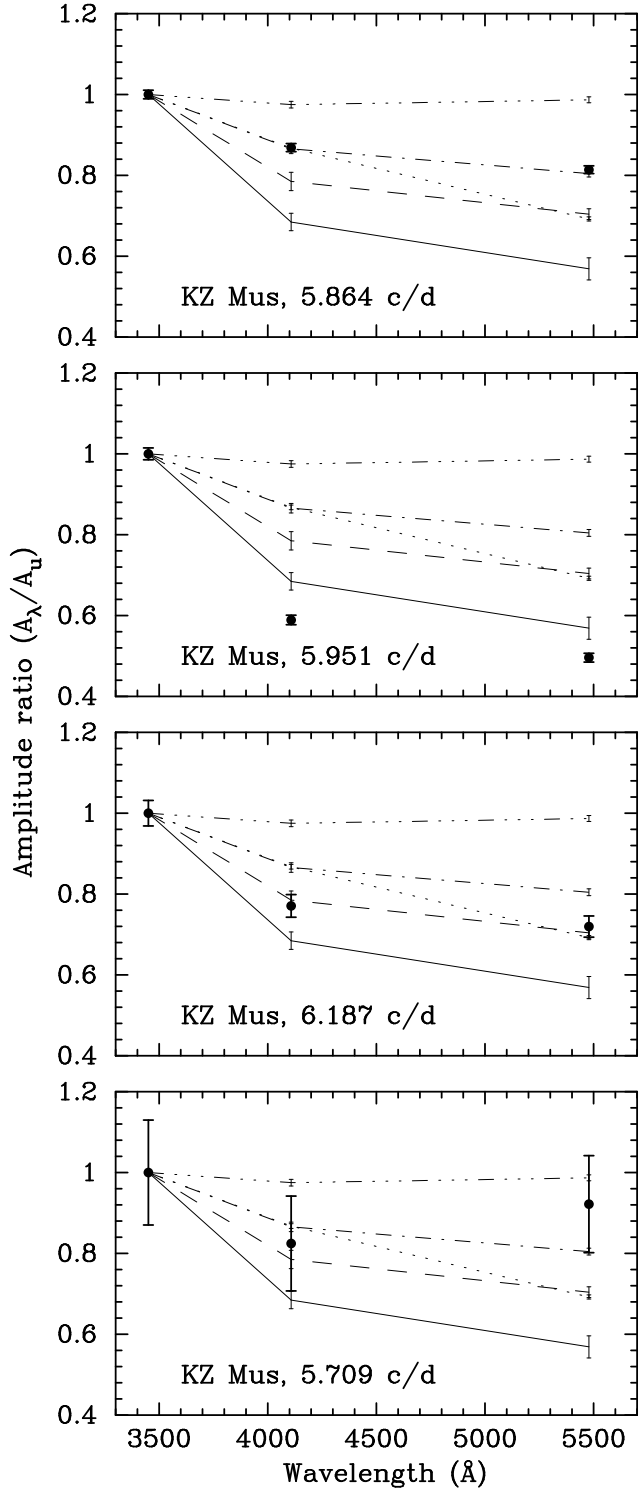}
\caption[]{Mode identifications for KZ Mus from a comparison of observed
and theoretical $uvy$ amplitude ratios, normalised to unity at $u$. The
filled circles with error bars are the observed amplitude ratios. The full
lines are theoretical predictions for radial modes, the dashed lines for
dipole modes, the dashed-dotted lines for quadrupole modes, the dotted 
lines for octupole modes, and the dashed-dot-dot-dotted lines are for 
$\ell=4$. The thin error bars denote the uncertainties in the theoretical 
amplitude ratios.}
\end{figure}

Turning to KZ Mus, one immediately sees that its strongest mode is
$\ell=2$. The second strongest mode is suggested to be radial, although
again the observed amplitude ratios indicate a steeper wavelength
dependence of the amplitudes than theoretically predicted. The third mode
of KZ Mus is a dipole, $\ell=1$, and the fourth is nonradial. These
results are in perfect agreement with the mode identifications by Handler
et al.\ (2003).

\section{Discussion and conclusions}

We have observed the three $\beta$ Cephei stars because we aimed at new 
and improved mode identifications and because we wanted to check whether 
these objects would justify a larger observational effort for 
asteroseismic purposes. We have reasonably succeeded in providing the mode 
identifications.

One point that appears problematic, however, is that the theoretically
predicted $u/v$ and $u/y$ amplitude ratios for radial modes are smaller
than the measured ones for $\beta$ CMa and KZ Mus. Whereas the reason for
this discrepancy may be incorrect differential colour extinction
coefficients for $\beta$ CMa that directly affect the measured amplitudes
of the 3.9994\cd mode, we cannot invoke this explanation for the 5.951\cd
mode of KZ Mus as its frequency is too different from an integer multiple
of one cycle per sidereal day. In addition, the theoretical and observed
$UvBV$ amplitudes derived by Handler et al.\ (2003) are in perfect
agreement for a radial mode in this star. We note that we have used the
same differential colour extinction correction for $\beta$ and 15 CMa,
which was derived from the 15 CMa data only. These two stars were measured
simultaneously, and have very similar colours, which is why this procedure
was soundest.

Returning to the mode identifications of the individual stars, the two
strongest modes of $\beta$ CMa are a quadrupole and likely a radial mode.  
Given the effective temperature, luminosity and thus mass inferred in
Sect.\ 4, we can compute the pulsation ``constant'' for such a radial
mode. The result of this calculation is $Q=0.035\pm0.009$\,d, indicating
that this mode may be the radial fundamental, although the possibility
that it is the first overtone cannot be ruled out this way.

We were also able to locate a radial mode in the pulsation spectrum of 15
CMa, whose pulsation constant would be $Q=0.027\pm0.007$\,d. This suggests
that we deal with the first radial overtone in this case, although the
fundamental mode or the second overtone can also not be excluded. We note,
however, that no convincing detection of a radial mode in a $\beta$ Cephei
star other than the fundamental has been made to date.

Three of the nonradial pulsation modes of 15 CMa are almost equally spaced
in frequency. Regrettably, we cannot be sure whether these modes are
$\ell=1$, 2 or 3. Independent of the $\ell$ value, we can still assume
that these modes do correspond to rotationally split components of the
same mode to check whether this leads to reasonable results.

Doing so, and further assuming that the observed splitting corresponds to
the surface rotational frequency of 15~CMa, we derive a rotation period of
9.35\,d, which translates into an equatorial rotational velocity of
$49\pm8$\,\kms. The measured $v \sin i$ of 15 CMa is 40\,\kms (Uesugi \&
Fukuda 1982), making it possible that the three nonradial modes of the
star indeed belong to the same $k$ and $\ell$.

For KZ Mus we confirmed the mode identifications by Handler et al.\
(2003). Unfortunately, the $\ell$ value of the weakest known independent
mode of the star can still not be uniquely constrained. Repeated
observations in the future may help to solve this problem as measurements
from different years can be analysed together, if homogeneous filter sets
(particularly bluewards of the Balmer jump) are used. If a mode
identification of $f_4$ can be derived, we believe it is likely that
theoretical asteroseismic investigations can be undertaken. Of course,
additional measurements in the future may also help to reveal modes beyond
our present detection limit (1.5 mmag).

To our regret, the asteroseismic prospects of $\beta$~CMa do not seem to
be very high. With only three modes photometrically detected and only two
identified with their $\ell$ value, a detailed exploration of its inner
structure seems unlikely. Perhaps this bright star is better studied by
means of spectroscopy, for instance as done by Briquet et al.\ (2006).

15 CMa seems more worthwhile for further study. We have proven the
existence of a radial mode, which is of tremendous importance for
restricting the star's location in the HR diagram. We have also found
evidence that the remaining three nonradial modes may form a rotationally
split triplet, which can be the starting point for an asteroseismic study.  
What still needs to be done is to prove that this is indeed a mode triplet
and to determine its $\ell$ value.

An independent clue towards the spherical degree of this possible triplet
may come from its frequency asymmetry due to the second-order effect of
rotation, which is different for different $\ell$. Following the
definition of the asymmetry of a frequency triplet $f_- < f_0 < f_+$ by
Dziembowski \& Jerzykiewicz (2003):

\begin{displaymath}
A_{obs}=f_- + f_+ - 2f_0,
\end{displaymath}

we obtain $A = -0.0077 \pm 0.0015$\cd from our measurements and $A =
-0.0004 \pm 0.0011$\cd from the archival data. These two values are
inconsistent (perhaps due to systematic frequency errors in the
single-site archival data) and can therefore not be interpreted; more
time-resolved measurements are needed.

An unusual feature in the light curves of 15 CMa is that the descending
branch of the strongest mode's pulsation is steeper than its rising
branch. In fact, even a ``stillstand'' phenomenon is indicated. This has
been found in only one $\beta$~Cephei star, BW Vul, to date (e.g.\ see
Sterken et al.\ 1986) and is usually interpreted as a shock phenomenon in
the atmosphere due to the vicious acceleration of the stellar material due
to pulsation (e.g.\ Fokin et al.\ 2004).

However, there are two differences between the pulsations of BW Vul and 15
CMa: BW Vul pulsates in a single radial mode, and with much higher
photometric amplitude than 15 CMa, whose mode showing the unusual light
curve form is nonradial. The small photometric amplitude can be explained
under the assumption that the strongest mode of 15 CMa is the $m=0$
component of a rotationally split multiplet. The possible range in the
inclination of the pulsation axis to the line of sight may lead to heavy
geometrical cancellation of such a mode (e.g.\ see Pesnell 1985). In other
words, this mode could have high intrinsic amplitude and its low
photometric amplitude is just a projection effect.

The unusual light curve form of this mode naturally gives rise to a
harmonic frequency with high relative amplitude compared to the mode
causing it. Combination frequencies with abnormally high amplitudes could
in some cases also be caused by resonant mode coupling (see Handler et
al.\ 2006 for a more detailed discussion), whereas other combination
frequencies may be simple light-curve distortions caused by a nonlinear
response of the flux to a sinusoidal displacement due to pulsation. The
sum frequencies we found in the amplitude spectra of KZ Mus seem to be
such ``normal'' combinations. Their relative amplitudes are in the same
range as those of the ``normal'' combinations of the $\beta$ Cephei stars
12 Lac (Handler et al.\ 2006) and $\nu$~Eri (Handler et al.\ 2004,
Jerzykiewicz et al. 2005), keeping in mind inclinational effects. Resonant
mode coupling is therefore not required to explain the combination
frequencies of KZ Mus.

Finally, we noticed the occurrence of slow intrinsic variability in our 
residual light curves for $\beta$ and 15 CMa. Such variations have
been reported for several other $\beta$ Cephei stars (see the discussion 
by Handler et al.\ 2006), and seem to be a fairly common phenomenon. 
Despite the existence of very extensive data sets, its physical cause 
remains unknown.

\section*{ACKNOWLEDGEMENTS}

This work has been supported by the Austrian Fonds zur F\"orderung der
wissenschaftlichen Forschung under grants R12-N02 and P18339-N08.

\bsp

\end{document}